\documentclass{PoS}
\usepackage{amsmath}
\pdfoutput=1

\title{Einstein-\ae ther gravity: a status report}

\ShortTitle{Einstein-\ae ther gravity}

\author{\speaker{Ted Jacobson}\\
        Center for Fundamental Physics\\
University of Maryland \\
College Park, MD 20742-4111, USA\\
        E-mail: \email{jacobson@umd.edu}}

\abstract{This paper reviews the theory,
phenomenology, and observational constraints on
the coupling parameters of Einstein-\ae ther gravity, i.e.\
General Relativity coupled
to a dynamical unit timelike vector field.
Several new remarks and a discussion of open
issues are included.}

\FullConference{From Quantum to Emergent Gravity: Theory and Phenomenology\\
        June 11-15 2007\\
        Trieste, Italy}

\begin{document}

\def\H{{\cal H}}
\def\ttheta{\tilde{\theta}}

\def\beq{\begin{equation}}
\def\eeq{\end{equation}}
\def\bea{\begin{eqnarray}}
\def\eea{\end{eqnarray}}
\def\ben{\begin{enumerate}}
\def\een{\end{enumerate}}
\def\la{\langle}
\def\ra{\rangle}
\def\a{\alpha}
\def\b{\beta}
\def\g{\gamma}\def\G{\Gamma}
\def\d{\delta}
\def\e{\epsilon}
\def\phi{\varphi}
\def\k{\kappa}
\def\l{\lambda}
\def\m{\mu}
\def\n{\nu}
\def\o{\omega}
\def\p{\pi}
\def\r{\rho}
\def\s{\sigma}
\def\z{\zeta}
\def\t{\tau}
\def\L{{\cal L}}
\def\S{\Sigma }
\def\gsim{\; \raisebox{-.8ex}{$\stackrel{\textstyle >}{\sim}$}\;}
\def\lsim{\; \raisebox{-.8ex}{$\stackrel{\textstyle <}{\sim}$}\;}
\def\gtrsim{\gsim}
\def\lessim{\lsim}
\def\loc{{\rm local}}
\def\vm{v_{\rm max}}
\def\bh{\bar{h}}
\def\del{\partial}
\def\nab{\nabla}
\def\half{{\textstyle{\frac{1}{2}}}}
\def\fourth{{\textstyle{\frac{1}{4}}}}

\section{Introduction}

There are reasons to suspect that the vacuum in
quantum gravity may determine a preferred rest frame
at the microscopic level. However, if such a frame
exists, it must be very effectively concealed from
view. Numerous observations severely limit the
possibility of Lorentz violating physics among the
standard model fields~\cite{Mattingly:2005re}. The
constraints on Lorentz violation in the gravitational
sector are generally far weaker.

To allow for gravitational Lorentz violation without
abandoning the framework of general relativity (GR),
the background tensor field(s) breaking the symmetry
must be dynamical. Einstein-\ae ther theory is of this
type. In addition to the spacetime metric tensor field
$g_{ab}$ it involves a dynamical, unit timelike vector
field $u^a$. Like the metric, and unlike other
classical fields, the unit vector cannot vanish
anywhere, so it breaks local Lorentz symmetry down to
a rotation subgroup. It defines a congruence of
timelike curves filling all of spacetime, like an
omnipresent fluid, and so has been dubbed the ``\ae
ther". This paper aims to provide a review of what has
been learned about this theory, the status of
observational constraints, and some currently open
issues.

A primary motivation for studying Einstien-\ae ther
theory---``\ae -theory " for short---is the quantum
gravity suspicion mentioned above. A secondary aim is
to develop a viable and reasonably natural foil
against which to compare gravitational observations,
in an era when numerous alternate gravity theories
have already been either ruled out or severely
constrained. A third source of interest is the
theoretical laboratory it offers for studying
diffeomorphism invariant physics with preferred frame
effects.

The action involving metric and \ae ther is highly constrained
by diffeomorphism invariance, locality,
and the unit constraint on $u^a$.
The only term with no derivatives is
the cosmological constant, there are no terms with
one derivative (other than a total divergence),
and there are five terms with two derivatives,
\beq S = -\frac{1}{16\pi G}\int \sqrt{-g}~ (R+K^{ab}_{mn} \nabla_a
u^m \nabla_b u^n)~d^{4}x. \label{action} \eeq
Here $R$ is the Ricci scalar, and the tensor
$K^{ab}_{mn}$ is defined by
\beq K^{ab}_{mn} = c_1 g^{ab}g_{mn}+c_2\d_m^a\d_n^b
+c_3\d_n^a\d_m^b +c_4u^au^bg_{mn}, \eeq
where the $c_i$ are dimensionless
coupling constants.  Like general relativity,
pure classical \ae -theory
defined by this action is
scale-free.
(The metric signature is $({+}{-}{-}{-})$,
the speed of light defined by the metric $g_{ab}$
is unity, and the \ae ther is taken to be dimensionless.)
The term $R_{ab}u^au^b$ can be expressed as the
difference of the $c_3$ and $c_2$ terms, up to a total
derivative, so is not independent.
In computations the unit timelike constraint on the
\ae ther is usually
imposed by adding a Lagrange multiplier term
$\l(g_{ab}u^au^b-1)$ to the action.
The covariant derivative
operator $\nabla_a$ involves derivatives of the metric
through the connection components, and the unit
vector is nowhere vanishing, hence
the terms quadratic in
$\nabla_a u^m$ are quadratic in derivatives of
both \ae ther and metric perturbations, so the
metric and \ae ther modes are coupled.
The field equations are written out in detail
in many of the references, beginning with
Ref.~\cite{Eling:2003rd}. It is noteworthy that
the \ae ther stress tensor includes second derivative
terms arising from the variation of the metric in the
connections. In all phenomenology work to date, it has
been assumed that the \ae ther is aligned
at large
scales with the rest frame of the microwave background
radiation.

The restriction to no more than two derivatives is
motivated by the standard precepts of effective field
theory~\cite{Burgess:2003jk}: higher derivatives would
be suppressed by one power of a small length scale for
each extra derivative. The natural size of the
coupling constants $c_i$ depends on unknown physics at
high energies. If the Planck scale is the only
relevant scale then the $c_i$ are naturally all of
order unity. If on the other hand there is an
additional scale or scales characterizing the Lorentz
violating physics, then the $c_i$ might be naturally
smaller and could differ from each other in order of
magnitude. Note that although the $c_4$ term in the
action is quartic in the \ae ther field, it makes a
quadratic contribution to the kinetic terms for metric
and \ae ther perturbations when expanded around a flat
background with a constant \ae ther of unit norm.

Einstein-\ae ther theory is similar to the
vector-tensor gravity theories studied long ago by
Will and Nordvedt~\cite{willnord}, but with the
crucial difference that the vector field is
constrained to have unit norm. This constraint
eliminates a wrong-sign kinetic term for the
length-stretching mode~\cite{Elliott:2005va}, hence
gives the theory a chance at being viable. An
equivalent theory using the tetrad formalism was first
studied by Gasperini~\cite{Gasperini}, and in the
above form it was introduced by Jacobson and
Mattingly~\cite{Jacobson:2000xp}. Related
vector-tensor gravity theories are that of Kostelecky
and Samuel~\cite{Kostelecky:1989jw} which corresponds
to the Maxwell-like special case of \ae -theory
($c_3=-c_1$, $c_2=c_4=0$), both with a fixed norm and
with a symmetry breaking potential for the vector, of
Gripaios~\cite{Gripaios:2004ms} which has all
two-derivative terms and a symmetry breaking potential
for the vector, and the generalized Einstein-\ae ther
theories of Zlosnik, Ferreira and
Starkman~\cite{Zlosnik:2006zu} and of
Zhao~\cite{Zhao:2007ce}, in which the \ae ther kinetic
terms of (\ref{action}) are replaced by functions
thereof. Kanno and Soda~\cite{Kanno:2006ty} introduced
dependence of the coupling parameters on a scalar
field that has its own dynamics, and studied
inflationary cosmology for one such model. Other
theories involving scalar fields in addition to a
vector field with a Maxwell-like action for the vector
are Bekenstein's TeVeS~\cite{Bekenstein:2004ne} and
Moffat's STVG~\cite{Moffat:2005si}.

\subsection{Matter couplings}
Lorentz-violating (LV) effects in the matter sector
are produced by couplings of the matter to the \ae
ther. (More complicated Lorentz symmetry breaking
patterns require LV extensions of general relativity
with other symmetry breaking fields~\cite{Bluhm:2004ep}.) Such
effects have been highly constrained by
observations~\cite{Mattingly:2005re},
so if they exist they are very weak. In this review I
will therefore assume that matter couples universally
to the metric $g_{ab}$. This assumption is motivated
by phenomenology, but goes against the
precepts of effective field theory since the Lorentz
violation in the gravitational sector would presumably
induce Lorentz violating terms in the matter action
via loop effects. Supersymmetry could conceivably provide a
natural suppression of these terms, as discussed briefly
in section~\ref{Frontiers}.

\subsection{Metric redefinitions}

When investigating aspects of the theory that do not
involve matter, it can be helpful to exploit the
metric redefinition $g'_{ab} = g_{ab}+(\z-1) u_a u_b$,
which ``stretches" the metric tensor in the \ae ther
direction by a positive factor $\z$.
(A negative factor would return a Euclidean signature metric.)
The action (\ref{action})
for ($g'_{ab}$, $u'^a = u^a/\sqrt{\z}$) takes the same
form as that for ($g_{ab}$, $u^{a}$), with new
coefficients $c'_i$. The relation between the $c'_i$
and $c_i$ was worked out in~\cite{Foster:2005ec}, and
is conveniently given in terms of certain combinations
with simple scaling behavior:
\bea
c'_{14} &=& c_{14}\nonumber\\
c'_{123} &=& \z c_{123}\nonumber\\
c'_{13}-1 &=& \z (c_{13}-1)\nonumber\\
c'_1-c'_3-1 &=& \z^{-1}(c_1-c_3-1). \label{redef} \eea
Note that in the absence of matter couplings
a one-parameter set of
Einstein-\ae ther theories is actually
pure vacuum GR in disguise~\cite{BarberoG.:2003qm,Foster:2005ec}:
the parameters
$c_i'$ all vanish if $c_{14}=c_{123}=2c_1-c_1^2+c_3^2=0$,
and $\z>0$ provided $c_{13}<1$.

\section{Newtonian and post-Newtonian limits}
\label{NPN}

In the weak-field, slow-motion limit \ae -theory
reduces to Newtonian gravity~\cite{Carroll:2004ai},
with a value of  Newton's constant $G_{\rm N}$ related
to the parameter $G$ in the action (\ref{action})  by
\beq G_{\rm N}=\frac{G}{1-c_{14}/2}, \label{GN}\eeq
where $c_{14}\equiv c_1+c_4$. (Similar notation is  used below for
other additive combinations of the $c_i$.) As long as  $c_{14}<2$,
the Newtonian limit is thus recovered. If $c_{14}>2$ gravity is
repulsive. This suggests the possibility of resolving
singularities by anti-gravity if the coupling coefficients $c_i$
were not constants~\cite{Gasperini}. Here I discuss only the
theory with constant coefficients.

All parameterized post-Newtonian (PPN)
parameters~\cite{willLR} of ae-theory except the
preferred frame parameters $\alpha_{1,2}$ agree with
those of GR for any choice of the $c_i$.  In
particular, the Eddington-Robertson-Schiff parameters
$\b$ and $\g$ are both unity~\cite{Eling:2003rd}, the
Whitehead parameter $\xi$
vanishes~\cite{Foster:2005dk}, and the five
energy-momentum conservation parameters $\a_3,
\zeta_{1,2,3,4}$ vanish simply because the theory is
derived from a lagrangian. The parameter $\a_2$ was
found for small $c_i$ in~\cite{Graesser:2005bg}, and
the exact values of $\a_1$ and $\alpha_2$ were found
in~\cite{Foster:2005dk}:
\bea
    \alpha_1&=& \frac{-8(c_3^2 + c_1c_4)}{2c_1 - c_1^2+c_3^2}\label{alpha1}\\
    \alpha_2&=&\frac{\a_1}{2}
    -\frac{(c_1+2c_3-c_4)(2c_1+3c_2+c_3+c_4)}{c_{123}(2-c_{14})}\label{alpha2} \eea
(This particular way of expressing $\a_2$ was given
in~\cite{Foster:2006az}.) In deriving the PPN parameters,
the $00$-component of the metric perturbation and the
$0$-component of the aether perturbation are expanded
to $O(2)$ in the Newtonian potential, while the
$0i$-components of the metric and $i$ components of
the aether are truncated at their lowest order,
$O(1.5)$.\footnote{The 0-component of the
aether is expanded explicitly only to $O(1)$
in Ref.~\cite{Foster:2005dk},
but the contribution of the $O(2)$ part is
implicitly incorporated by use of the identity (A.26).}

Observations currently impose the strong constraints
$\a_1 \lesssim 10^{-4}$ and $\a_2\lesssim 4\times
10^{-7}$~\cite{willLR}. Since \ae -theory has four
free parameters $c_i$, we may set $\alpha_{1,2}$
exactly to zero by imposing the
conditions~\cite{Foster:2005dk}
\begin{eqnarray}
 c_2&=&(-2c_1^2-c_1c_3 + c_3^2)/3c_1 \label{zeroalphac2}\\
 c_4&=&-c_3^2/c_1 \label{zeroalphac4}.
\end{eqnarray}
 With
(\ref{zeroalphac2}, \ref{zeroalphac4}) satisfied, {\it all} the PPN
parameters of \ae -theory are equivalent to those of GR.

The
parameters $\a_{1,2}$ can also be set to zero by imposing
$c_{13}=c_{14}=0$, but this case is pathological, as discussed in
section \ref{special}.  Also discussed there is the case $c_{123}=0$,
for which $\a_2$ can diverge. Another special case occurs
when $2c_1-c_1^2 -c_3^2=0$, for which $\a_1$ diverges unless
also $c_3^2+c_1c_4=0$ in which case it is indeterminate.
Together these imply $c_{14}=2$, so
Newton's constant (\ref{GN}) diverges in this case.

\section{Homogeneous isotropic cosmology}

Assuming spatial homogeneity and isotropy, $u^a$
necessarily coincides with the 4-velocity of the
isotropic observers, and automatically satisfies its
field equation. The \ae ther stress tensor is
constructed purely form the spacetime geometry and is
identically conserved. It is just a certain
combination of the Einstein tensor and the stress
tensor of a perfect fluid with energy density
proportional to the inverse square of the scale
factor, like the curvature term in the Friedman
equation~\cite{Mattingly:2001yd,Carroll:2004ai},
\beq T^{\mbox{\scriptsize \ae ther}}_{ab}= -\frac{c_{13}+3c_2}{2}
\Bigl[G_{ab}-\textstyle{\frac16}{}^{(3)}R(g_{ab}+2u_au_b)\Bigr].
\eeq
The latter contribution plays no important
cosmological role since the spatial curvature is
small, while the former renormalizes the gravitational
constant appearing in the Friedman equation,
yielding~\cite{Carroll:2004ai}
\beq G_{\rm cosmo}=\frac{G}{1+(c_{13}+3c_2)/2}. \eeq
Since $G_{\rm cosmo}$ is not the same as $G_{\rm N}$
the expansion rate of the universe differs from what
would have been expected in GR with the same matter
content. The ratio is constrained by the observed
primordial ${}^4$He abundance to satisfy $|G_{\rm
cosmo}/G_{\rm N} - 1|\lessim1/8$~\cite{Carroll:2004ai}. When
the PPN parameters $\a_{1,2}$ are set to zero by
(\ref{zeroalphac2}, \ref{zeroalphac4}), it turns out
that $G_{\rm cosmo}=G_{\rm N}$, so this
nucleosynthesis constraint is automatically
satisfied~\cite{Foster:2005dk}.

\section{Linearized wave modes}
\label{modes}

When linearized about a flat metric and constant \ae
ther, \ae -theory possesses five massless modes for
each wave vector: two spin-2, two spin-1, and one
spin-0 mode. The squared speeds of these modes
relative to the \ae ther rest frame are given
by~\cite{Jacobson:2004ts}
\bea\label{speeds}
\mbox{spin-2}\qquad&&1/(1-c_{13})\label{s2}\\
\mbox{spin-1}\qquad&&(2c_1- c_1^2+
c_3^2)/2c_{14}(1-c_{13})\label{s1}\\
\mbox{spin-0}\qquad&&c_{123}(2-c_{14})/c_{14}(1-c_{13})(2+c_{13}+3c_2)\label{s0}
\eea
The corresponding polarization tensors were found in
one gauge in Ref.\ \cite{Jacobson:2004ts} and in
another gauge in Ref.\ \cite{Foster:2006az}. The
former gauge choice for the metric and \ae ther
perturbations $h_{ab}$ and $v^a$ is $h_{0i}=0$ and
$v^i\!,\!{}_{i}=0$. In this gauge the nonzero components of
the polarization for waves in the 3-direction are
given by
\bea\label{pols}
\mbox{spin-2}\qquad&&h_{12},\quad h_{11}=-h_{22}\label{pol2}\\
\mbox{spin-1}\qquad&&v_I,\quad h_{3I}= \left[
{2c_{14}c_{13}^2}/{(2c_1- c_1^2+
c_3^2)}\right]^{1/2}\, v_I\label{pol1}\\
\mbox{spin-0}\qquad&& v_0,\quad h_{00}=-2v_0,\quad
h_{11}=h_{22}=-c_{14}v_0,\quad
h_{33}=[2c_{14}(1+c_2)/c_{123}]\, v_0,
\label{pol0}\eea
where the index $I$ takes only the transverse values
$1,2$.

The energy of the spin-2 modes is always positive,
while for the spin-1 modes it has the sign of $(2c_1
-c_1^2 +c_3^2) /(1-c_{13})$, and for the spin-0 modes
it has the sign of
$c_{14}(2-c_{14})$~\cite{Eling:2005zq,Foster:2006az}.
(These reduce to the results of Ref.\
\cite{Lim:2004js} in the decoupling limit where
gravity is turned off.) Under the metric redefinition
$g_{ab}\rightarrow g_{ab}+(\z-1)u_au_b$, (\ref{redef})
can be used to verify that the squared speeds all
scale as $1/\z$ as expected, and the signs of the
energy are invariant.

Several constraints are imposed by stability
conditions on the linearized modes. The squared speeds
correspond to (frequency/wavenumber)${}^2$, so must be
non-negative to avoid imaginary frequency
instabilities. They must moreover be greater than or
equal to unity, to avoid the existence of vacuum
\v{C}erenkov radiation by
matter~\cite{Elliott:2005va}. (The existence of ultra
high energy cosmic rays, which propagate near the
metric light cone, requires that the metric and \ae
ther modes propagate at no less than an extremely
small amount below the metric speed of light.) And the
mode energy densities should be positive, to avoid
dynamical instabilities.

With the $\a_{1,2}=0$ conditions
(\ref{zeroalphac2}, \ref{zeroalphac4}) imposed, the squared speeds
expressed in terms of $c_\pm\equiv c_1\pm c_3$ are
\bea\label{zaspeeds}
\mbox{spin-2}\qquad&&1/(1-c_+)\\
\mbox{spin-1}\qquad&&(c_++c_--c_+c_-)(c_++c_-)/4c_+c_-(1-c_+)
\label{s1za}\\
\mbox{spin-0}\qquad&&c_+/3c_-(1-c_+), \eea
the sign of the spin-1 mode energy is that of
$(c_++c_--c_+c_-)/(1-c_+)$, and $c_{14}=2c_+c_-/(c_++c-)$. The
conditions that the squared spin-2 and spin-0 mode speeds be
greater than or equal to unity then restrict $c_\pm$ to the region
\bea\label{superluminal}
0&\le& c_+\le1\label{sl1}\\
0&\le& c_-\le c_+/3(1-c_+).\label{sl2} \eea
Some algebra shows that the inequalities $0\le
c_+\le1$ and $0\le c_-$ also ensure that the squared
spin-1 mode speed (\ref{s1za}) is also greater than or
equal to unity. They also imply $0\le c_{14}<2$, hence
the condition for attractive gravity mentioned in
section \ref{NPN} need not be separately imposed. Also
the energies of the spin-0 and spin-1 modes are then
positive. That is, when $\a_{1,2}=0$ all of the
requirements for all of the modes are met and gravity
is attractive if and only if  and $c_\pm$ are
restricted by the inequalities (\ref{sl1},
\ref{sl2})~\cite{Foster:2005dk}. The allowed range is
the shaded region in Fig.~1.
\begin{figure}[htb]\label{c+c-plane}
\begin{center}\hspace{-1.2cm}
\includegraphics[width=4in]{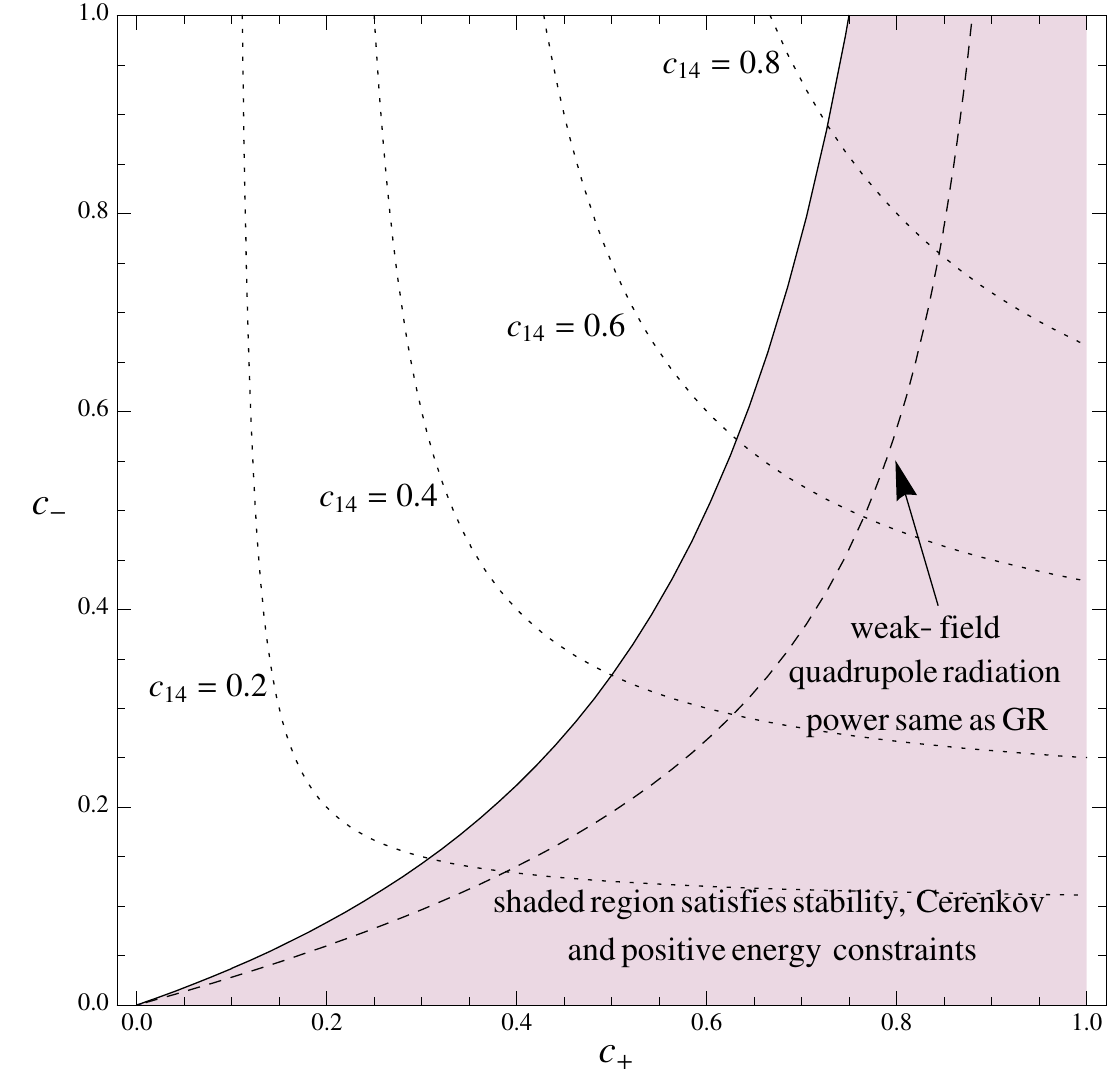}
\end{center}
\caption {The shaded region
satisfies the stability, \v{C}erenkov, and positive energy
constraints (\protect\ref{sl1}, \protect\ref{sl2}) on $c_\pm$ after
having imposed the vanishing
of the PPN parameters $\a_{1,2}$ by the conditions
(\protect\ref{zeroalphac2}, \protect\ref{zeroalphac4}).
The dashed curve is the locus where the
net radiation power from a source with
weak self-field matches that of GR. The four dotted curves
are contours of fixed $c_{14}$.}
\end{figure}

Interestingly, if the mode speeds are instead required to be {\it
less} than unity (sub-luminal), then the spin-1 and spin-0 energy
densities are negative. Hence not only the \v{C}erenkov
constraint, but also energy positivity (together with
$\a_{1,2}=0$) requires mode speeds greater than unity.

\section{Primordial perturbations}

The behavior of primordial cosmological perturbations
in ae-theory has been studied in
Refs. \cite{Lim:2004js,Li:2007vz}. The coupling to the
\ae ther modifies the spin-2 perturbations in two
ways: it changes their propagation speed, as already
discussed above for perturbations of flat space, and
it changes their amplitude, since that is determined
by the ``bare" $G$ in the action (\ref{action}) rather
than by the measured Newton constant $G_{\rm N}$. The
result is that, given the same $G_{\rm N}$,  and
assuming the PPN parameters $\a_{1,2}$ vanish, the
power in these perturbations differs from that in GR
by the factor $(1-c_{14}/2)(1-c_{13})^{1/2}$. When the
stability constraints (\ref{sl1}, \ref{sl2}) are
satisfied this factor is smaller than unity, hence
these spin-2 perturbations are even more difficult to
detect than in GR.

Perturbations of the \ae ther can be decomposed into
spin-0 and spin-1 modes, both of which decay by
themselves as the universe expands. In the presence of
a scalar inflaton, however, the spin-0 \ae ther mode
is sourced and no longer decays away. For single
scalar field slow-roll inflation, it turns out that
the power spectrum is generally modified, but it is
unchanged when the PPN parameters vanish.

The late time evolution of the primordial
perturbations is modified, since anisotropic photon
and neutrino stresses in the radiation dominated epoch
source the spin-1 mode. This leads to modified matter
and CMB spectra~\cite{Li:2007vz}. The effect is rather
small however, and is degenerate with matter-galaxy
bias and with neutrino masses.

Primordial perturbations are thus difficult to distinguish in
ae-theory and GR.

\section{Radiation damping and strong self-field effects}

Two-body dynamics and radiation in ae-theory have been
analyzed for purely weak fields in
Ref.~\cite{Foster:2006az}, and including strong self
gravity effects in Ref.~\cite{Foster:2007gr}. In
general the radiation includes not only spin-2 waves
sourced by a varying quadrupole moment, but also
spin-1 waves sourced by quadrupole and dipole moments
and spin-0 waves sourced by quadrupole, dipole, and
monopole moments. As seen in section \ref{modes}, the
waves of different spin propagate in general at
different speeds relative to the \ae ther rest frame.
Direct observation of all of these waves is in
principle possible with existing gravitational wave
antennas~\cite{willLR}. With waves of sufficient
amplitude from multiple sources, and a network of
detectors, the \ae ther rest frame and the wave speeds
(\ref{s2})-(\ref{s0}) and polarizations 
(\ref{pol2})-(\ref{pol0}) could all be measured.

\subsection{Weak fields}
If the fields are weak everywhere (including inside
the radiating bodies),
 and
the PPN parameters $\a_{1,2}$ vanish, then the dipole
source vanishes and the monopole source is uncoupled
to the fields. In this regime therefore all radiation
is sourced by the quadrupole, as in GR. The net power
radiated in spin-0, spin-1, and spin-2 modes in this
case is given by $(G_{\rm N} {\mathcal
A}/5)\dddot{Q}_{ij}^2$, where $Q_{ij}$ is the
quadrupole moment and
 ${\mathcal A}={\mathcal A}[c_i]$ is a
 function of the coupling parameters $c_i$
that reduces to unity in the case of GR. Agreement
with the damping rate of GR (confirmed to $\sim 0.1\%$
in binary pulsar systems~\cite{willLR}) can be
achieved by imposing the condition ${\mathcal
A}[c_i]=1$, which is consistent with the constraints
(\ref{sl1}, \ref{sl2}), as shown by the dashed line in
Fig.~1.

\subsection{Strong field effects}
For compact sources with strong internal fields such
as neutron stars or black holes, terms of higher order
in $Gm/d$ (where $m$ and $d$ are the mass and
characteristic size of the source) can play an
important role in the dynamics and radiation. Since
background fields and radiation fields vary little
over compact source dimensions, these strong field
effects can be handled using an ``effective source"
dynamics specified by a worldline action
integral~\cite{Foster:2007gr}
\beq S=-m_0\int d\t\; [1
+\s(v^au_a-1)+\s'(v^au_a-1)^2+\dots],
\label{effectiveaction} \eeq
where $v^a$ is the 4-velocity of the body, $u_a$ is
the local background value of the \ae ther, and $\s$
and $\s'$  are a constants characterizing the body,
called a ``sensitivity parameters" or just
``sensitivities". A similar technique was introduced
by Eardley~\cite{Eardley} for scalar-tensor gravity,
in which the first sensitivity measures the dependence
of the action on the local background value of the
scalar field. The ellipses in (\ref{effectiveaction})
stand for all other invariants that can be formed from
the particle worldline and the background fields. All
of these are of higher order in the velocity relative
to the \ae ther or involve more derivatives and are
hence suppressed. Among other things, the
sensitivities determine the dependence of energy on
velocity, $E=m_0 + \half (1+\s)m_0v^2
+\textstyle\frac38(1+\s-\s')m_0v^4+\dots$. (This is
obtained by expressing the action
(\ref{effectiveaction}) in the form $\int dt (T-V)$
where $V=m_0$, and reading off $E=T+V$.) At lowest
order in binding energy $\Omega$, $\s$ is given by
\beq \s= (\a_1-\textstyle{\frac{2}{3}}\a_2)(\Omega/m)
+ \mathcal{O}\Bigl(f[c_i](G_{\rm N}m/d)^2\Bigr), \eeq
where $f[c_i]$ scales as $c_i$ for
$c_i\ll1$.\footnote{This corrects an error in version
1 of Ref.\ \cite{Foster:2007gr}, where $\s$ is said to
scale as $c_i^2$. (Also the a prefactor $c_{14}$ in
Eqn. (70) should be deleted.) As a result of this
correction, the likely constraints on $c_i$ are an
order of magnitude stronger, as stated
here~\cite{bzf-pc}.} Note that the leading term
vanishes when $\a_{1,2}$ vanish. The higher order
terms in $\s$ have not yet been computed, nor have
even the lowest order terms in $\s'$ (which will also
scale as $c_i$). They presumably differ for black
holes and stars, and will depend on the equation of
state of the star.

Nonzero sensitivities lead to a number of
phenomena that are constrained by observations,
including dependence of dynamics on the ambient
\ae ther vector,
modified post-Newtonian
two-body dynamics, modified quadrupole sourced
radiation, and both monopole and dipole sourced
radiation. When $\a_{1,2}=0$, all of these constraints
are met provided the sensitivities are less than $\sim
0.001$, which will certainly be the case if
$c_i\lesssim 0.01$~\cite{Foster:2007gr}. To be more
precise would require knowing the actual dependence of
the sensitivities on the $c_i$. It is conceivable that
the constraints will be much weaker.

To determine the strong field constraints just
mentioned, it is necessary to make some assumption
about the speed $V$ of the observed systems with
respect to the background \ae ther frame.  It was
argued in Ref.~\cite{Foster:2007gr} that neglect of
terms involving this speed is justified provided
$V\lesssim 10^{-2}$, which is easily satisfied for any
known proper motion relative to the rest frame of the
microwave background radiation.

\section{Spherically symmetric stars and black holes}

Unlike GR, \ae -theory has a spherically symmetric
mode, corresponding to radial tilting of the \ae ther, and
there is a three-parameter family of
spherically symmetric static vacuum solutions~\cite{Eling:2006df}.
If asymptotic flatness is imposed and the mass
fixed, there remains a one-parameter
family~\cite{Eling:2003rd,Eling:2006df}, whereas
GR has the unique Schwarzschild solution.
As explained below, the evidence suggests that
only two solutions in this one-parameter family
are astrophysically relevant,
at least provided the couplings are not too large.

The total energy of an asymptotically flat
solution of \ae -theory
receives a contribution from the asymptotic
$1/r$ falloff of the
\ae ther~\cite{Eling:2005zq,Foster:2005fr}.
The \ae ther normalization condition locks this
contribution to that of the metric, so the energy can be
expressed in terms of the metric alone.
It turns out
to be just $E=r_0/G_{\rm N}$, where
$r_0$ is the coefficient of the $1/r$ term in
the $g^{rr}$ inverse metric component\footnote{This
is the asymptotic Misner-Sharp energy~\cite{Hayward:1994bu},
covariantly defined by
$r_0=\lim_{r\rightarrow\infty} (r/2)(1-g^{ab}r_{,a}r_{,b})$,
where $r$ is the area radius.}
and $G_{\rm N}$ is Newton's constant (\ref{GN}).
That is, it agrees with the Newtonian mass in the
Newtonian limit, which is inevitable given that the
theory has a Newtonian limit.  This is how the mass of the
solutions discussed below is determined.

\subsection{Static vacuum \ae ther}
The solution outside a static star is the unique
vacuum solution for a given mass in which the \ae ther
is aligned with the Killing
vector~\cite{Eling:2006df}. This ``static \ae ther"
vacuum solution depends on the $c_i$ only through the
combination $c_{14}$, and was found analytically (up
to inversion of a transcendental
equation)~\cite{Eling:2006df}. It has a globally
timelike Killing vector, and is asymptotically flat.
The spatial slices have a minimal area 2-sphere,
inside of which the 2-spheres flare out to infinite
area at a singularity on a would-be Killing horizon
with vanishing surface gravity. The affine parameter
distance to the singularity is finite along radial
null geodesics. (The minimal two sphere is a property
of the vacuum solution only; it never occurs in the
presence of a matter source.) Although this static
wormhole indicates the presence of a negative
effective energy density in the field equation (i.e.
the Einstein tensor has this property), all solutions
in this family have positive total mass. In
Ref.~\cite{Seifert:2007fr} it was found that this
static \ae ther solution is stable to linear
perturbations under the same conditions as for flat
spacetime, with the exception of the case $c_{123}=0$.

\subsection{Neutron stars}

The solution inside a fluid star has been found by
numerical integration, both for constant
density~\cite{Eling:2006df} and for realistic neutron
star equations of state~\cite{Eling:2007xh}. The
maximum masses for neutron stars range from about 6 to
15\% smaller than in GR when $c_{14}=1$, depending on
the equation of state. The corresponding surface
redshifts can be as much as 10\% larger than in GR for
the same mass. Measurements of high gravitational
masses or precise surface redshifts thus have the
potential to yield strong joint constraints on
$c_{14}$ and the equation of state.

Sufficiently compact stars can have around them an
innermost stable circular orbit (ISCO), which occurs
in the Schwarzscild solution of GR at the radius
$6G_{\rm N}M$. The ISCO in the static \ae ther
solution lies at a radius larger by approximately the
factor $(1+0.03\, c_{14})$, with orbital frequency
lower by the factor $(1-0.04\, c_{14})$. These small
differences would likely be difficult to measure in
practice.

\subsection{Black holes}
For black holes the \ae ther cannot be aligned with
the Killing vector, since the latter is not timelike
on and inside the horizon. Instead, the \ae ther is at
rest at spatial infinity and flows inward at finite
radii. The condition of regularity at the spin-0
horizon (where the outgoing spin-0 waves propagate at
constant radius) selects a unique solution from the
one-parameter family of spherical stationary solutions
for a given mass~\cite{Eling:2006ec}. This is because
the coefficient of the second radial derivative term
in a field equation vanishes at the spin-0 horizon,
which forces the second derivative to blow up unless
the rest of the terms also vanish. Generically this
regularity condition does not hold, but the solution
can be tuned to satisfy it. When a black hole forms
from collapse of matter, the spin-0 horizon develops
in a nonsingular region of spacetime, where the
evolution should be regular. This motivated the
conjecture that collapse produces a black hole with
nonsingular spin-0 horizon, which has been confirmed
for some particular examples in numerical simulations
of collapse of a scalar field~\cite{Garfinkle:2007bk}.

The black holes with nonsingular spin-0 horizons are
rather close to Schwarzschild outside the horizon for
a wide range of couplings; for instance, the ISCO
radius differs by a  factor $(1 + 0.043 c_1 + 0.061
c_1^2)$, in the case with $c_3=c_4=0$ and $c_2$ fixed
so that the spin-0 speed is unity~\cite{cte-pc}. (This
expansion is accurate at least when $c_1\le0.5$. No
solution with regular spin-0 horizon exists in this
case when $c_1 \gsim 0.8$.) Inside the horizon the
solutions differ more, but like Schwarzschild they
contain a spacelike singularity. Black hole solutions
with singular spin-0 horizons have been studied in
Ref.\ \cite{Tamaki:2007kz}. These solutions can differ
much more outside the horizon. Quasi-normal modes of
black holes in \ae -theory have been investigated in
Refs.\ \cite{Konoplya:2006rv}.

\section{Special values of $c_i$}
\label{special}

It would be interesting if there existed a symmetry of
the action restricting the values of the coupling
parameters $c_i$, but none has been identified. Values
for which the PPN parameters $\a_{1,2}$ vanish
evidently lead to more symmetry in the weak field
limit, but this does not appear to extend to a
symmetry of the full theory.  A microscopic theory
might predict a particular relation or hierarchy
between the coefficients, but no such theory is
in hand. Hence any prior imposition
of restrictions would be unjustified.
Nevertheless, we
may ask from a purely phenomenological perspective
whether some simplifying restrictions could be
compatible with observations. The answer is negative.
It appears that, unless {\it all} of the parameters
are much smaller than unity, then
in order to be compatible with all of
the theoretical and observational constraints (leaving
aside the radiation reaction and strong self-field effects),
most likely {\it none} of the $c_i$, nor $c_{13}$,
$c_{14}$, or $c_{123}$ can vanish. We now discuss
these special cases in some detail.

\begin{itemize}

\item The first case to be examined in
detail~\cite{Kostelecky:1989jw,Jacobson:2000xp} was
$c_{13}=c_2=c_4=0$, i.e.\ the ``Maxwell action" (with
the unit constraint on the vector). The PPN result for
$\a_2$ (\ref{alpha2}) is infinite in this case, and
the spin-0 mode speed is zero. The perturbation series
used in the PPN analysis is thus evidently not
applicable. Independently of that, however, other
problems with this case have been identified, such as
the formation of shock
discontinuities~\cite{Jacobson:2000xp,Clayton:2001vy}
and a possibly related
instability~\cite{Seifert:2007fr}.
This case is equivalent to Maxwell theory
in a special gauge if
it is further restricted to the sector in which the
Lagrange multiplier field vanishes (see
~\cite{Nambu68, Jacobson:2000xp,
Chkareuli:2007da,Bluhm:2007bd} and references therein).
However, this equivalence
holds only if one abandons the notion that the vector
is itself physical. The shock discontinuities and
instabilities of the vector can then be regarded as
gauge artifacts, but the theory is no longer
that of a Lorentz-violating vector field.

\item The mode speeds (\ref{speeds}) are all equal to unity
if
$c_{13}=c_4=0$ and $c_2=c_1/(1-2c_1)$.
In this case the PPN parameters
become $\a_1=-4c_1$ and $\a_2=0$,
so the observational
constraint on $\a_1$
imposes the stringent condition
$|c_1|\lesssim 2.5\times 10^{-5}$, and the other
parameters are then similarly small.
Positivity of the spin-1 and spin-0 mode energies
imposes $0<c_1<1$. With small positive $c_1$, the
remaining constraints can be satisfied as well.

\item The case $c_1=c_3=0$ is evidently not
covered by the existing PPN analysis, since $\a_1$
(\ref{alpha1}) is indeterminate and the spin-1 mode
speed is zero. (If $c_2=0$ then also the spin-0 speed
vanishes. If $c_4=0$ then the spin-1 speed is
indeterminate and the spin-0 speed is infinite, but
the energy of both modes is zero.) It
appears that the post-Newtonian order of some field
quantities must be modified in this case, which could
be interesting to examine ab initio.

\item Assuming that (i) $\a_{1,2}=0$,   (ii) the
speed constraints
(\ref{sl1}, \ref{sl2}) are satisfied, (iii) the spin-1 mode
speed (\ref{s1}) is not infinite with finite energy,
and  (iv) putting aside the case
$c_1=c_3=0$ which is not covered by existing
PPN analyses, the only viable
case in which {\it any} of $c_i$,
$c_{13}$, $c_{14}$, or $c_{123}$ vanishes is
the special case
$c_3=c_4=2c_1+3c_2=0$, with $2/3<c_1<1$.
This large value of $c_1$
is probably inconsistent with the strong field constraints from
orbital binaries, but as mentioned above those are not yet
precisely known because the sensitivity parameters have not yet
been computed for neutron stars, so this case is not yet ruled
out. All the other special cases contradict one or more of
the conditions (i)-(iv), which can be seen as follows.
We halt when a case is reduced to a previous case.

\begin{itemize}

\item $c_1=0$:  $\a_1=0$ then implies also $c_3=0$. As stated above,
the case $c_1=c_3=0$ requires a new analysis.

\item $c_{13}=0$: The speed constraint (\ref{sl2}) then implies
also $c_-=0$, so $c_1=c_3=0$.

\item  $c_{14}=0$: $\alpha_1=0$  then implies $c_3^2= c_1^2 $, so either
$c_+ = 0$ or $c_- = 0$. In either case, the spin-1 speed diverges
unless $c_1=0$.

\item $c_{123}=0$: The denominator of the second term in
$\alpha_2$ vanishes (\ref{alpha2}), so either $\a_2$ diverges
or,  if the numerator also vanishes, it is indeterminate.
When $\a_1=0$, $\a_2$ becomes
\beq
\a_2=-\frac{c_{13}^2(3c_1c_{123}-c_{13}^2)}{c_1^2c_{123}(2-c_{14})},
\eeq
so the numerator vanishes in this case only if $c_{13}=0$.
(This also implies $c_{14}=0$.)

\item $c_2=0$: $\a_{1,2}=0$ implies (\ref{zeroalphac2}),
which can be written as
\beq
c_2  = -c_+(c_+ + 3c_-)/6c_1.
\label{za2'}
\eeq
Thus either $c_+ = 0$ (covered above) or
$c_+ + 3c_- = 0$. In the latter case,  either $c_+$ or
$c_-$ is negative, violating the speed constraints.

\item $c_3=0$: $\alpha_1=0$ implies $c_1=0$ (covered above) or
$c_4=0$. In the latter case,   (\ref{zeroalphac2}) or (\ref{za2'})
then implies that $2c_1 + 3c_2=0$.

\item $c_4=0$:  $\alpha_1=0$ then implies $c_3=0$.

\end{itemize}

\end{itemize}

\section{Frontiers}
\label{Frontiers}

A number of questions remain open regarding phenomenology
of \ae -theory, as well as more fundamental issues. In this
section some of these are briefly outlined.

\begin{itemize}

\item The most pressing issue is to compute the
sensitivity parameters for neutron stars and black
holes, so that strong field effects on radiation
damping rates, waveforms, and post-Newtonian dynamics
can be evaluated. To determine the sensitivities for a
given compact object one might compute (presumably
numerically) the exact solution for the object moving
at some velocity with respect to the asymptotic
aether, and match e.g.\ multipole moments of the
asymptotic fields to those of the corresponding
solution derived from the worldline action
(\ref{effectiveaction}). It may be adequate to solve
for linear perturbations of the exact, static
solutions, rather than finding the exact moving
solutions, but this may not be any more convenient.

\item Spherical collapse of a star to a black hole
would be accompanied by a pulse of spin-0
radiation, since the \ae ther must adjust from the
static configuration (parallel to the Killing vector)
to the black hole configuration (flowing across the
horizon). This radiation may be much stronger than
that of ordinary gravitational waves arising from
asphericity of supernova core collapse. Moreover, it
could in principle be measured by gravitational wave
detectors, as seen from the polarization components
(\ref{pol0}).

\item  Rapidly rotating black holes,
unlike the non-rotating ones, might turn out to be
very different from the Kerr metrics of GR. If so,
this could provide an interesting observational
signature to detect or constrain. It may be that the
easiest way to determine such solutions would be by
numerical simulation of axisymmetric collapse.

\item Although the linear perturbations all
have positive energy for
coupling parameters within in a certain range,
the total, nonlinear energy has not been
shown to be always positive in this range.
(The positive energy theorem of GR does not
apply since the
\ae ther stress tensor does not satisfy the
dominant energy condition.)
Even a spherically
symmetric result in this regard would be useful. The
question of energy positivity was addressed in
Ref.~\cite{Clayton:2001vy}, but attention was
restricted there to the case where the coupling to
gravity is neglected, and only some special cases were
examined.

\item Supersymmetry could potentially allow for a naturally
suppressed Lorentz violation in the matter
sector~ \cite{GrootNibbelink:2004za}.
To include gravity in such a
scheme would require a Lorentz violating supergravity.
Can \ae -theory be supersymmetrized?
The argument~\cite{Deser:1977hu} that supergravity
has positive energy and therefore so must GR would
presumably extend to
\ae -theory. Therefore, supersymmetrizability is
likely only possible at most for
those parameter ranges for which
\ae -theory has positive energy.

\item
Black hole thermodynamics remains a puzzle  in \ae
-theory. It has not been possible to identify the
entropy and a corresponding ``first law" of black hole
mechanics, despite the existence of an identity
relating mass and angular momentum variations to a
horizon integral~\cite{Foster:2005fr}. The difficulty
is related to the fact that the \ae ther necessarily
diverges at the bifurcation surface, so the methods
usually employed~\cite{Wald:1993nt} are not applicable.
One might expect that the problem is related to the fact
that, in a theory with multiple limiting speeds, no
unique horizon is selected. However, even for the
cases mentioned in section \ref{special} for which the
mode speeds are all equal to unity, the entropy has
not yet been identified.

A further important issue is the status of the
generalized second law of thermodynamics. It has been
argued that in a theory with more than one limiting
speed, both quantum Hawking
radiation~\cite{Dubovsky:2006vk} and classical energy
extraction~\cite{Eling:2007qd} methods can be
exploited to violate the generalized second law of
thermodynamics. This suggests that there may be some
deep problem with such theories, including ae-theory
(with unequal mode speeds).

\item Lorentz-violating matter fields can be consistently
coupled to gravity using \ae -theory. In this way
gravitational back-reaction of such fields is well
defined. This has been studied in a cosmological
setting~\cite{Jacobson:2000gw}, and the issue of
renormalization of the corresponding quantum stress tensor has
recently been seriously tackled~\cite{Lopez Nacir:2007jx}.
Counter-terms renormalizing the \ae -theory action are
generally required. Such analyses should enable the
quantum back reaction to cosmological expansion as
well as to Hawking radiation to be studied.

\item Quantization of ae-theory may be instructive to
investigate. The dynamical unit timelike vector field
determines a time flow with respect to which evolution
can be referred.  This ``material reference system"
could perhaps be
employed~\cite{Brown:1994py,Brown:1995fj} to address
the ``problem of
time"~\cite{Isham:1992ms,Kuchar:1999br} in canonical
quantum gravity.

\item Vacuum ae-theory in 1+1 dimensions might provide
a useful simplified model for studying quantum
effects. The classical theory has been
investigated~\cite{Eling:2006xg}, where the full
solution space was mapped out. The solutions include
de Sitter and anti-de Sitter spacetimes, with a
uniformly accelerated \ae ther that is invariant under
a two-dimensional subgroup of SO(2,1) generated by a
boost and a null rotation. (The \ae ther becomes
infinitely stretched on certain Killing horizons in
these solutions.) The only other solution is a
peculiar one with singularities and no Killing
vectors.

\end{itemize}

\section{Conclusion}

Einstein-\ae ther theory provides a theoretical
laboratory in which gravitational effects of possible
Lorentz violation can be meaningfully studied without
abandoning the generally covariant framework of
general relativity. It is striking that \ae -theory
can match observations to the degree that it can.
While of course starting with four free parameters
$c_{1,2,3,4}$ makes this easier, the freedom to choose
these parameters might have already been used up in
setting the PPN parameters to agree with those of GR.
Instead, only two parameters need be fixed at that
stage, leaving $c_1$ and $c_3$ free. Then, despite
possessing three distinct types of linearized wave
modes, all positive energy, stability and vacuum
\v{C}erenkov requirements on the modes are met within
a large common region of the $(c_1,c_3)$ space. Within
this same parameter space, the dynamics of the
cosmological scale factor and perturbations differ
little from GR, and non-rotating neutron star and
black hole solutions are quite close to those of GR
(but may be distinguishable with future observations).

Additional constraints arise from the observed
radiation damping rate in binaries. For systems with
weak self-fields, a constraint of order $10^{-3}$
would be imposed on one combination of the two
parameters $c_{1}$ and $c_3$. Current observations
involve pulsars which possess strong self-fields.
Presuming that the sensitivity parameters for neutron
stars turn out to have the expected magnitude, such
binaries will exhibit
effects constraining both the parameters to be less
than around $10^{-2}$, in addition to the stronger
radiation damping constraint on one combination. It
may turn out that observations of multiple systems
with different combinations of sensitivities will
constrain both parameters separately at the $10^{-3}$
level.

\section*{Acknowledgments}
I am grateful to C.T.\ Eling, B.Z.\ Foster, B.\ Li, E.\ Lim,
D.\ Mattingly, I.\ Rothstein
and T. Sotiriou
for helpful discussions and comments on drafts of
this article. This work was supported by NSF grant
PHY-0601800.

\end{document}